\documentclass[twocolumn,showpacs,preprintnumbers,amsmath,amssymb]{revtex4}

\usepackage{graphicx}
\usepackage{dcolumn}
\usepackage{bm}

\begin{document}

\title{Diluted Graphene Antiferromagnet}

\author{L. Brey$^1$, H. A. Fertig$^{2,3}$, and S. Das Sarma$^{4}$}
\affiliation{1. Instituto de Ciencia de Materiales de Madrid (CSIC),
Cantoblanco, 28049 Madrid, Spain\\
2. Department of Physics, Indiana University, Bloomington, IN 47405\\
3. Department of Physics, Technion, Haifa 32000, Israel\\
4. Department of Physics, University of Maryland, College Park, MD 20742}

\date{\today}

\begin{abstract}
We study RKKY interactions between local magnetic moments for both
doped and undoped graphene.  We find in both cases that the
interactions are primarily ferromagnetic for moments on the same
sublattice, and antiferromagnetic for moments on opposite
sublattices. This suggests that at sufficiently low temperatures dilute
magnetic moments embedded in graphene can order into a state
analogous to that of a dilute antiferromagnet.  We find that in
the undoped case one expects no net magnetic moment, and
demonstrate numerically that this effect generalizes to ribbons
where the magnetic response is strongest at the edge, suggesting
the possibility of an unusual spin-transfer device.  For doped
graphene we find that moments at definite lattice sites interact
over longer distances than those placed in interstitial sites
of the lattice ($1/R^2$ vs. $1/R^3$) because the former support a
Kohn anomaly that is suppressed in the latter due to the absence
of backscattering.
\end{abstract}
\maketitle

{\it Introduction--} Graphene, a two-dimensional honeycomb network of carbon
atoms, has recently become a subject of intense interest.  The development
of practical fabrication techniques for single graphene sheets \cite{Novoselov_2004}
has allowed experimental study of this system, confirming its two-dimensional
Dirac spectrum in quantum Hall studies and revealing many unique
properties \cite{Novoselov_2005, Zhang_2005}.  In this work we study
response functions of graphene in the non-interacting limit, focusing
on its consequences for magnetic moments which may be embedded in the
system (RKKY interactions).  A number of studies of magnetic moments
in graphene have identified a tendency toward antiferromagnetic
\cite{Fujita_1996,Son_2006,Pisani_2007,Yazyev_2007} or ferromagnetic \cite{Vozmediano_2005}
correlations.  These correlations are usually attributed to exchange
interactions or other many-body effects.  In what follows, we
demonstrate that such effects arise even for non-interacting electrons
in graphene, and that they are a result of the chirality
of the electron states for doped graphene \cite{Ando_2005},
and of the vanishing density of states at the Fermi energy for undoped systems.
A non-interacting model of graphene may be justified by its small, density-independent
effective $r_s$ parameter, as well as studies suggesting that
Fermi liquid theory should work well in
physically realizable situations
\cite{Dassarma_2007,Polini_2007,Gonzalez_1994,Dahal_2006}.

Exchange coupling between local magnetic moments and conduction
electrons in metals leads to an effective (RKKY) coupling
\cite{Kittel_1968}  among the local moments which
oscillates with distance with wavevector $2k_F$ ($k_F=$ Fermi
wavevector), and an amplitude that decays as $1/R^2$ in two
dimensions, with $R$ the separation between impurities. For doped
graphene, we shall demonstrate similar behavior, with an important
qualitative difference: the sign of the interaction depends on
whether the two local moments couple to the honeycomb network on
sites of the same sublattice or different ones, and when summed
over both sublattices at a fixed distance, the $1/R^2$
contribution to the RKKY coupling is cancelled, leaving behind a
residue that falls off as $1/R^3$. Interestingly, analogous
studies of the linear response to perturbations that do not
distinguish between A and B sublattice sites also result in a
$1/R^3$ behavior \cite{Vozmediano_2005, cheianov_2006,
wunsch_2006}. We will show that the $1/R^3$ behavior -- and the
absence of $1/R^2$ behavior in density response functions -- is a
direct result of the chiral nature of electrons in graphene.

For undoped graphene ($k_F \rightarrow 0$)
we find the RKKY coupling behaves as $\sim 1/R^3$ at large
distances, again with equal magnitudes, that are
{\it ferromagnetic} when
the impurities are on the same sublattice, and {\it antiferromagnetic}
when on opposite sublattices.
This behavior is also connected to that of the full density
response, and reflects the vanishing density
of states of graphene at the Fermi points.  This behavior
also dominates in doped graphene for distances $R \lesssim 1/k_F$
where the coupling is greatest in magnitude.
Because of this we
expect at zero temperature the system will tend to
order, with moments oriented in opposite directions for the
two sublattices.  The state is thus analogous to an ordered state
of a dilute antiferromagnet.  Analogous behavior has been noted
in zigzag graphene ribbons \cite{Son_2006} with equal and opposite spin
accumulating near the edges, on opposite sublattices.  While
this effect has been attributed to complicated exchange
interactions \cite{Son_2006}, we present results of simple tight-binding
calculations demonstrating that this physics occurs even
without interactions, and is a consequence of the unusual
nature of the single particle states in graphene.

{\it Hamiltonian, Wavefunction, and RKKY Interaction --}
The simplest description of graphene is a tight-binding model
representing electrons in $\pi_z$ orbitals of the carbon atoms,
which can hop with matrix element $t$ between nearest neighbor
sites, which are always on opposite sublattices for the honeycomb lattice.
The energy states of such a model may be straightforwardly computed \cite{Ando_2005},
and one finds that the spectrum possesses particle-hole symmetry,
with a zero energy surface consisting of six points at
corners of the Brillouin zone, only 2 of which are inequivalent
due to symmetry. When undoped the Fermi surface of graphene passes
through these points, which are denoted by $K$ and $K'$.
At long wavelengths,
the wavefunctions near each of these points can be described by
two component spinor envelope functions
$[\phi^{A(\prime)}({\bf r}), \phi^{B(\prime)}({\bf r})]$, the
entries of which are proportional to the amplitude for the electron to
be present at unit cell located at ${\bf r}$ on sublattice
$A$ or $B$.  The wavefunctions may be regarded as possessing
a quantum number $\tau=K,K'$ denoting which Dirac point they
reside near.  The Hamiltonian near such a point is approximately
\begin{equation}
H^{(\prime)}= \pm v_F \left ( \begin{array}{cc}
0 & -i\partial_x \mp \partial_y \nonumber \\
-i\partial_x \pm \partial_y  & 0  \nonumber \\
\end{array} \right ) \, \, \, \, \, \,
\end{equation}
with the upper (lower) sign denoting the Hamiltonian for states
near the $K$ ($K'$) point, and $v_F=\sqrt{3}t/2$.  These
Hamiltonians have eigenenergies $\epsilon _{{\bf k},s}=s v_F |k|$,
and associated eigenstates $\psi^{(\prime)}_{{\bf k},s}=(e^{\mp i \theta_{\bf k}},\pm s)$,
where again the upper (lower) sign denotes the solution for the
$K$ ($K'$) valley, $s=\pm 1$,
and $\theta_{\bf k}=\arctan (k_x/k_y)$.

Consider local spin degrees of freedom
${\bf S}_{\mu}({\bf R}_1)$ and ${\bf S}_{\nu}({\bf R}_2)$ weakly
coupled to electrons in graphene by an exchange interaction $J$
at positions at or near sites in sublattices $\mu$ and $\nu$.
In perturbation theory \cite{Fischer_1975, Beal_1987}
the induced interaction between the
spins has the form $H_{\mu\nu}=J_{RKKY}^{\mu,\nu}{\bf S}_{\mu} \cdot {\bf S}_{\nu}$, where
$J_{RKKY}^{\mu,\nu}=-J^2\chi^0_{\mu,\nu}({\bf R}_1-{\bf R}_2)$, and
$\chi^0_{\mu,\nu}$ is the Fourier transform of
\begin{equation}
\chi ^0 _{\mu,\nu} (q)  = - g_v {1 \over N}\sum _{s,s',{\bf k}} \frac
{f(\epsilon _{{\bf k},s})-f(\epsilon _{{\bf k}+ {\bf q},s'})}
{\epsilon _{{\bf k},s}-\epsilon _{{\bf k}+ {\bf q},s'}}
F_{s,s'}^{\mu,\nu}({\bf k},{\bf q}) \, \, . \label{chimunu}
\end{equation}
Here $g_v=2$ is the degeneracy due to the valley index,
$N$ is the number of unit cells in
the system, $f$ is the Fermi function, and
$F_{s,s'}^{\mu,\nu}({\bf k},{\bf q})$ is a factor arising
from the matrix element of the spinors associated with
the single particle states, which in general depend on
the angles $\theta_{\bf k}$ and $\theta_{{\bf k}+{\bf q}}$
\cite{com_F}.

{\it Site-Symmetric Moments --} When the local moments are
located at the centers of the hexagons in the honeycomb network,
it becomes appropriate to replace $F_{s,s'}^{\mu,\nu}$
Eq. \ref{chimunu} with a sum,
$F_{s,s'}=\sum_{\mu,\nu}F_{s,s'}^{\mu,\nu}
=\frac {1}{2} ( 1+ s s' \cos{\Delta \theta _{{\bf
k}+ {\bf q}}})$, where $\Delta \theta _{{\bf k}+ {\bf q}}$ is the angle formed by
the vectors ${\bf k}$ and ${\bf k}+{\bf q}$. The resulting
$\chi_0$ is then identical to the standard density-density
response function, which may be computed straightforwardly
\cite{wunsch_2006, Ando_2006, Hwang_2006}, with a result
that may be expressed conveniently in the form
$\chi ^0 (q, \mu)  = \chi ^0  (q, \mu =0) + \Delta \chi ^0 (q,\mu)$
with $\mu=v_Fk_F$ the chemical potential (assumed positive),
$\chi ^0  (q, \mu =0) = \frac {g_v q}{16 v_F}$, and
\begin{eqnarray}
&~&\Delta\chi ^0 (q,\mu)
 =   \frac {g_v k_F }{2 \pi v_F} \left (1- \frac {\pi}{4}
\frac{q}{2 k_F} \right ) \Theta (2k_F-q)  \nonumber \\ & & + \frac
{g_v k_F }{2 \pi v_F} \left ( 1 -\frac {1}{2} \sqrt{ 1 - \left
(\frac {2k_F}{q} \right )^2 } -\frac {1}{2} \frac {q}{2k_F} \arcsin
\frac {2 k_F}{q} \right ) \nonumber \\
&~&\times \Theta (q-2k_F). \label{pi0(qm)}
\end{eqnarray}

Several comments are in order.
(1) In spite of the presence of step functions $\Theta$ in this expression,
its first derivative with respect to $q$
is continuous at $q=2k_F$, in sharp contrast with
the situation for a normal two dimensional electron gas (2DEG).
The discontinuity in the 2DEG arises from a singularity in the
integrand in Eq. \ref{chimunu} (with $F=1$ for a 2DEG) when
$\epsilon_{\bf k}=\epsilon_{{\bf k}+{\bf q}}$ and $q=2k_F$ --
the Kohn anomaly \cite{Kohn_1957}.
For graphene, $F_{s,s'}$ vanishes precisely where the singularity
would otherwise occur, removing the discontinuity in the slope.
This behavior is a direct result of the chirality of electrons
in graphene and the resulting absence of backscattering that
it entails \cite{Ando_2005}.
(2) For undoped graphene the response
vanishes at $q=0$.  We can understand this as follows.
The $q=0$ response may be understood as arising from
a shift in the chemical potential, plus more generally a part coming from changes
in the single particle wavefunctions.
However, the
{\it total} charge of the system cannot shift due to changes in
the single particle wavefunctions, in accordance with
the Friedel sum rule \cite{mahan_book}.  Moreover,
in undoped graphene the response from a differential chemical potential shift
vanishes because the density of states at
the Fermi energy is zero.  Thus there can be no net $q=0$ response.   (3) The vanishing of
$\chi_0$ at $q=0$ means that the total population of either spin
flavor cannot be changed by a perturbation in undoped graphene, even if the perturbation
is different for the two spin directions -- as would be the case
for a (possibly inhomogeneous) Zeeman coupling.
This result is consistent with the observation
that graphene ribbons can have an inhomogeneous spin configuration
but net spin zero \cite{Son_2006}.

{\it Site-Specific Moments -- } Local moments can in many circumstances
be more strongly coupled to a specific site in the honeycomb network,
which lies on a definite sublattice.  One can also consider situations
in which the moment is a substitutional impurity, or is an induced
moment due to a vacancy in the lattice \cite{Vozmediano_2005, Yazyev_2007}.
In such cases the coupling among moments has the form
$J_{RKKY}^{\mu,\nu} \propto \chi_{\mu,\nu}^0$, and $\chi_{\mu,\nu}^0$
is given by
Eq. \ref{chimunu} with
$
F_{s,s'} ^{A,A} ({\bf k},{\bf q})= \frac {1}{4} \, \, \, $
for impurities on the same sublattice, and
$
F_{s,s'} ^{A,B} ({\bf k},{\bf q})= \frac {1}{4} s s' e^ { i
\Delta\theta _{{\bf k}+ {\bf q}}}\, \, \,
$ for impurities on opposite sublattices.  We first consider the
case of impurities on the same sublattice.  Decomposing the
response function as
$
\chi ^0 _{A,A} (q)
= \chi ^0  _{A,A} (q, \mu =0) + \Delta \chi ^0 _{A,A} (q,\mu),
$
the first term, corresponding to undoped graphene, may be shown to have the form
\begin{equation}
\chi^0 _{A,A} (q, \mu =0)
=\frac{1}{2} \frac {g_v } {4 \pi v_F} (\Lambda- \frac {\pi
}{8} q )
\label{chi0aa}
\end{equation}
where $\Lambda \sim \pi/a_0$ is the momentum cutoff.  The contribution
due to doping may also be evaluated, and has the form
\begin{eqnarray}
& &\Delta\chi ^0 _{A,A} (q,\mu)
 =   \frac {g_v} {64 v_F } q
\Theta (2k_F-q) \nonumber \\
 &+&  \frac {g_v q}
{32\pi v_F }  \left [\arcsin(\frac {2 k_F}{q}) - \frac {2k_F}{q} \sqrt {1 -
\frac {4 k_F^2}{q^2}}  \right ] \Theta
(q-2k_F) \label{chi0(qm)}
\end{eqnarray}
In Eq. \ref{chi0(qm)} the derivative
is discontinuous at $q=2k_F$: the chiral overlap factor
$F^{AA}$ does not vanish in this case, and one obtains
a Kohn anomaly analogous to that of the standard 2DEG.
This has important consequences for RKKY coupling in
real space, which is proportional to the Fourier
transform of Eqs. \ref{chi0aa} and \ref{chi0(qm)}.
For the first of these we find
\begin{equation}
J_{RKKY} ^{AA} (R,\mu=0) \propto -\chi(R,\mu=0) =
-\frac{\pi}{32} \frac {g_v}{v _F} \frac {1}{R^3} \label{jR0mu0}
\end{equation}
so that in undoped graphene, moments are {\it ferromagnetically} coupled
when they are on the same sublattice.  The correction due to
doping, $\Delta J_{RKKY}^{AA}(R,\mu) \propto -\Delta\chi^0_{A,A}(R,\mu)$
can be computed in the asymptotic limit ($k_FR \gg 1$), with the result
\begin{equation}
\Delta\chi^0_{A,A}(R,\mu) \simeq
 \frac{g_v k_F}{4v_FR^2}
\sin (2k_F R)+  \frac
{g_v} {8v_FR^3}  (\cos (2k_F R)-1).
\end{equation}
A
comparison with numerical integration shows that this asymptotic expression
works quite well for $k_FR > 0.35$.
The oscillating term proportional to $1/R^2$ is present because the
Kohn anomaly is not suppressed in the relevant response function.
A similar behavior was found recently for Friedel oscillations,
where the way in which the perturbation breaks the lattice symmetry
determines whether they fall off as $1/R^2$ or $1/R^3$ \cite{cheianov_2006}.
While this $1/R^2$ behavior is similar to that of the
standard 2DEG, it nevertheless differs from the 2DEG in having
a density dependent amplitude \cite{Beal_1987}.

For moments on opposite sublattices, we can easily compute the coupling
by noting that $F^{A,A}_{s,s'}+F^{A,B}_{s,s'}=F_{s,s'}/2$.  It immediately
follows that
\begin{equation}
\chi ^0 _{A,B} (q,\mu)
=  - \chi ^0 _{A,A} (q,\mu) +\frac{1}{2} \Delta \chi ^0
(q,\mu).
\end{equation}
We thus see that the tendency towards ferromagnetic coupling for
moments within a distance $R \lesssim 1/k_F$ for impurities on the
same sublattice translates into an antiferromagnetic coupling
for impurities on opposite sublattices \cite{com_op}.  Moreover because the
coupling is strongest for short distances, we expect this to
result in a tendency towards antiferromagnetic order at low temperatures
when the moment density $n_i$ satisfies $k_F/\sqrt{\pi n_i} \lesssim 1$.
The low temperature state is analogous to that of a dilute antiferromagnet
since the moment locations are random in such models.  A special feature
of the graphene system, however, is that the coupling among the moments
can be manipulated via the electron density, which in turn may be controlled
by a gate \cite{Novoselov_2004}.  In particular, added electrons shorten
the distance over which the RKKY coupling has a well-defined (i.e., non-oscillating)
sign, so that the antiferromagnetic order may be suppressed via doping.
It is interesting to note that analogous, albeit simpler, behavior
(e.g., ferromagnetic rather than antiferromagnetic ordering) is believed to
occur in dilute magnetic semiconductors \cite{Brey_2003, Priour_2006}.
The physics associated with the chirality of the single-particle
states, as well as the vanishing density of states at the Fermi energy when
undoped, give graphene a richer phenomenology.

\begin{figure}
  \includegraphics[clip,width=9cm]{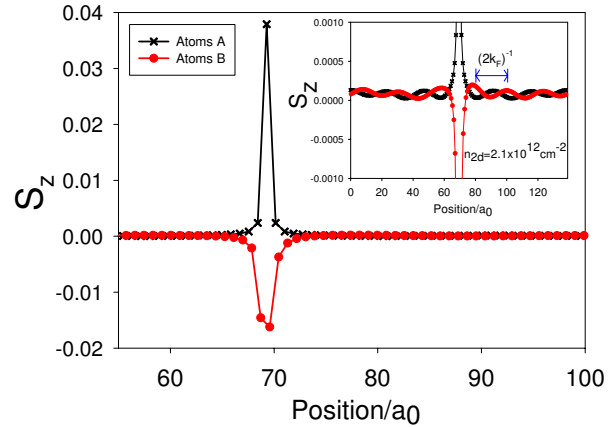}
  \caption{($Color$ $online$) Spin density as a function of position
  for a Zeeman field $E_z=t/10$ along line of sites in sublattice A for a ribbon
  geometry.  Main panel essentially identical for doped and undoped graphene.
  Inset: Blowup for undoped graphene illustrating RKKY oscillations.
  }
   \label{Figure1}
\end{figure}

{\it Numerical Investigations --} To test these results we have
performed numerical tight-binding calculations on graphene
ribbons.  We first consider a ribbon with zigzag edges,
with a Zeeman coupling ($E_z=t/10$) along a line of sites all on one
sublattice (A) near the center of the ribbon.  This type of perturbation models
a line of frozen spins.
Figure \ref{Figure1}
shows the results for the induced spin density, with A sites
shown in red and B sites in black.  The main panel is essentially
identical for both the doped and undoped cases.  In the doped
case one can see oscillations of wavevector $2k_F$
falling off slowly with distance, which are out-of-phase
for the two sublattices.  Moreover, the total induced spin
{\it vanishes} for the undoped case.  These properties
are in precise agreement with our expectation that summing over
sublattices leads to a cancellation of the RKKY oscillations
due to the absence of backscattering in graphene, and a vanishing net response
as $q \rightarrow 0$  due to the vanishing
density of states for undoped graphene.

\begin{figure}
  \includegraphics[clip,width=9cm]{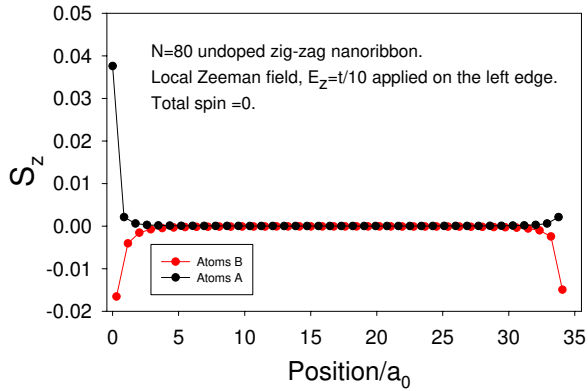}
  \caption{($Color$ $online$) Spin density as a function of position
  for a Zeeman field $E_z=t/10$ along the left edge of an
  undoped zigzag graphene ribbon,
  of width $N=80$ atoms along an armchair chain.  Spins induced at both
  edges, with a profile such the $S_z$ summed over all sites {\it vanishes}.
  Note that the perturbation applied to the left edge induces a strong
  response at the right edge, of net spin opposite that induced by the
  Zeeman field.
  }
   \label{Figure2}
\end{figure}

We also find an interesting result when
the perturbation is applied at one
of the edges (Fig. \ref{Figure2}).  Applying
a Zeeman field at a single zigzag edge in undoped graphene induces spin in
{\it both} edges, but in such a way that there is no induced total
spin for undoped graphene.  This is interesting because the spin state is communicated
across the width of the sample even though there is no spin polarization
in the bulk.  Thus the tendency for undoped graphene to compensate
an induced local spin due to a local Zeeman field survives the inclusion
of edge effects, which in the zigzag case induces a non-vanishing
density of states at zero energy \cite{Fujita_1996} for sufficiently
wide ribbons \cite{Brey_2006}.
We find results similar to those of Fig. \ref{Figure2}
for doped graphene zigzag ribbons with edge
Zeeman fields, with two differences: there are $2k_F$
oscillations in the spin density of small magnitude
as one moves in from the edge, and a small net spin
is induced.  We note that analogous spin configurations have been
predicted to {\it spontaneously} form in ribbons when exchange
interactions are important \cite{Son_2006}; our calculations demonstrate
that such interactions are not needed to induce the tendency towards
spin compensation.  It is interesting to speculate that this effect
might be utilized as a spin transfer device.

In summary, we have studied RKKY interactions among magnetic moments
in graphene using a linear response approach.  Our calculations show
a strong qualitative difference between moments that couple symmetrically
to the sublattices of the graphene honeycomb network and ones that
couple to specific sublattices, with the latter showing more pronounced
effects.  Doped graphene in particular supports oscillations due to
the Kohn anomaly only in the latter case.  The sum of intra- and intersublattice
responses was shown to vanish in the long wavelength limit in undoped
graphene, leading to RKKY interactions of opposite sign for the
two sublattices.  Within mean-field
theory, impurities
coupled via these interactions should form a low temperature state
analogous to that of a dilute antiferromagnet.  Tight-binding calculations
confirm the presence of the $2k_F$ oscillations for doped graphene,
and the tendency of opposite sublattices to have compensating spins.

After this work was completed, we became aware of related work \cite{Saremi_2007}
by S. Saremi
on undoped graphene, which also concludes that the sign of RKKY interactions
depends on whether moments are located on the same or opposite sublattices.

This work was
supported by MAT2006-03741 (Spain) (LB), by the NSF through Grant
No. DMR-0454699 (HAF), and by the US-ONR (SDS).


\begin{thebibliography}{27}
\bibitem[{com({\natexlab{a}})}]{com_F}
\bibinfo{note}{For the relevant matrix elements the $F$ factors do not depend
  on valley index, although in general they can. We do not consider the
  possibility of intervalley scattering, which introduces rapidly oscillating
  terms which vanish when averaged over position, and do not affect the net
  magnetization induced by RKKY interactions.}
\end{thebibliography}

\end{document}